\documentclass[oneside,british]{mem}
\usepackage[T1]{fontenc}
\usepackage[latin9]{inputenc}
\usepackage{amssymb}
\usepackage{graphicx}
\usepackage[authoryear]{natbib}

\makeatletter
\usepackage{natbib}
\usepackage{txfonts}
\usepackage{balance}
\usepackage{graphicx}
\usepackage[breaklinks,dvipdfm]{hyperref}\idline{1}{1}

\makeatother

\usepackage{babel}
\begin{document}

\title{Massive Binary Stars and Self-Enrichment of Globular Clusters}

\author{Robert~G.~Izzard\inst{1}, Selma~E.~de~Mink%
\thanks{Hubble fellow.%
}\inst{2,3}, Onno~R.~Pols\inst{4}, Norbert~Langer\inst{1}, Hugues~Sana\inst{5}
and Alex~de~Koter\inst{5}}

\institute{Argelander Institut f\"ur Astronomy, Universit\"at Bonn, Germany.
\and Space Telescope Science Institute, Baltimore, Maryland, U.S.A.\and 
Johns Hopkins University, Baltimore, Maryland, U.S.A. \and Department
of Astrophysics/IMAPP, Radboud University Nijmegen, The Netherlands.
\and Astronomical Institute Anton Pannekoek, University of Amsterdam,
The Netherlands.\vspace{-3mm}}

\abstract{~Globular clusters contain many stars with surface abundance patterns
indicating contributions from hydrogen burning products, as seen in
the anti-correlated elemental abundances of e.g. sodium and oxygen,
and magnesium and aluminium. Multiple generations of stars can explain
this phenomenon, with the second generation forming from a mixture
of pristine gas and ejecta from the first generation. We show that
massive binary stars may be a source of much of the material that
makes this second generation of stars. Mass transfer in binaries is
often non-conservative and the ejected matter moves slowly enough
that it can remain inside a globular cluster and remain available
for subsequent star formation. Recent studies show that there are
more short-period massive binaries than previously thought, hence
also more stars that interact and eject nuclear-processed material.}

\authorrunning{R.G.~Izzard~et~al.}

\titlerunning{Massive Binaries and GC Self-Enrichment}

\maketitle

\section{The mass budget and enrichment candidates}

The abundance correlations and helium enrichment observed in globular
cluster stars imply that proton-burning reactions are responsible
\citep[and many contributions to this volume]{2007A&A...470..179P}.
Hot hydrogen burning makes helium, nitrogen and aluminium, while destroying
oxygen, carbon and magnesium, as required in models of self-enrichment
in globular clusters. However, the number of stars in a second, or
further, generation is often similar to or exceeds the number in the
first generation \citep{2009A&A...505..117C}, and the amount of nuclear-processed
material currently in their atmospheres is similar to, or larger than,
that present in the atmospheres of the first stellar generation. It
is not clear how so much nuclear-processed mass can end up in the
second generation of stars. Four main channels have been investigated
to date:

1. \emph{Massive Asymptotic Giant Branch (AGB) stars} are the canonically
accepted prime candidates for self-enrichment \citep{2001ApJ...550L..65V}.
During their thermally-pulsing AGB (TPAGB) phase, hot-bottom burning
effectively cycles the whole stellar envelope through a hot hydrogen
burning shell. A star of mass $4\mathrm{\, M_{\odot}}\lesssim M\lesssim10\mathrm{\, M_{\odot}}$
ejects about $(M-1)\mathrm{\, M_{\odot}}$ of nuclear-processed material,
which is about 10\% of the mass of the whole stellar generation. This
does not take into account binary interaction which reduces the nuclear-processed
TPAGB mass yield \citep{2004MmSAI..75..754I} while allowing for significant
helium enrichment \citep{2012A&A...543A...4V}.

2. \emph{Rapidly rotating massive stars} also eject hydrogen-burned
material if they spin fast enough \citep{2007A&A...464.1029D}. Rotational
mixing transports material from the hot stellar core to the surface
where it is ejected if the star exceeds its critical rotation rate.
This is predicted to happen in some stars \citep{2012arXiv1211.3742D}
although the number of rapidly rotating stars is such that only 3\%
of the mass of all massive stars is ejected in this manner \citep{2009A&A...507L...1D}.

3. \emph{Stellar mergers} in dense cores of globular clusters may
also contribute to the reservoir of nuclear processed material \citep{2009A&A...497..255G}
although this channel probably does not contribute enough mass to
make the second generation of stars \citep{2010MNRAS.407..277S}.

4. \emph{Massive binary stars} are another source of nuclear processed
material, as we explore in the following.

\section{Massive binary stars}

While there is some doubt about whether most stars are in multiple
stellar systems, we can be sure that most stars with masses exceeding
about $2\mathrm{\, M_{\odot}}$ live with a companion star \citep{2007A&A...474...77K,2010ApJS..190....1R,2012ApJS..203...30F}.
Just as importantly, the latest estimate of the O-type binary-period
distribution in young, open clusters shows that more of them are close\emph{,
}i.e. liable to interact by mass transfer, than previously thought
\citep{2012Sci...337..444S}. Only about $29\%$ of O-type stars evolve
as single stars: the rest either have their envelope stripped ($33\%$),
merge ($24\%$) or accrete mass ($14\%$).

\begin{figure*}
\begin{centering}
\includegraphics[scale=0.5]{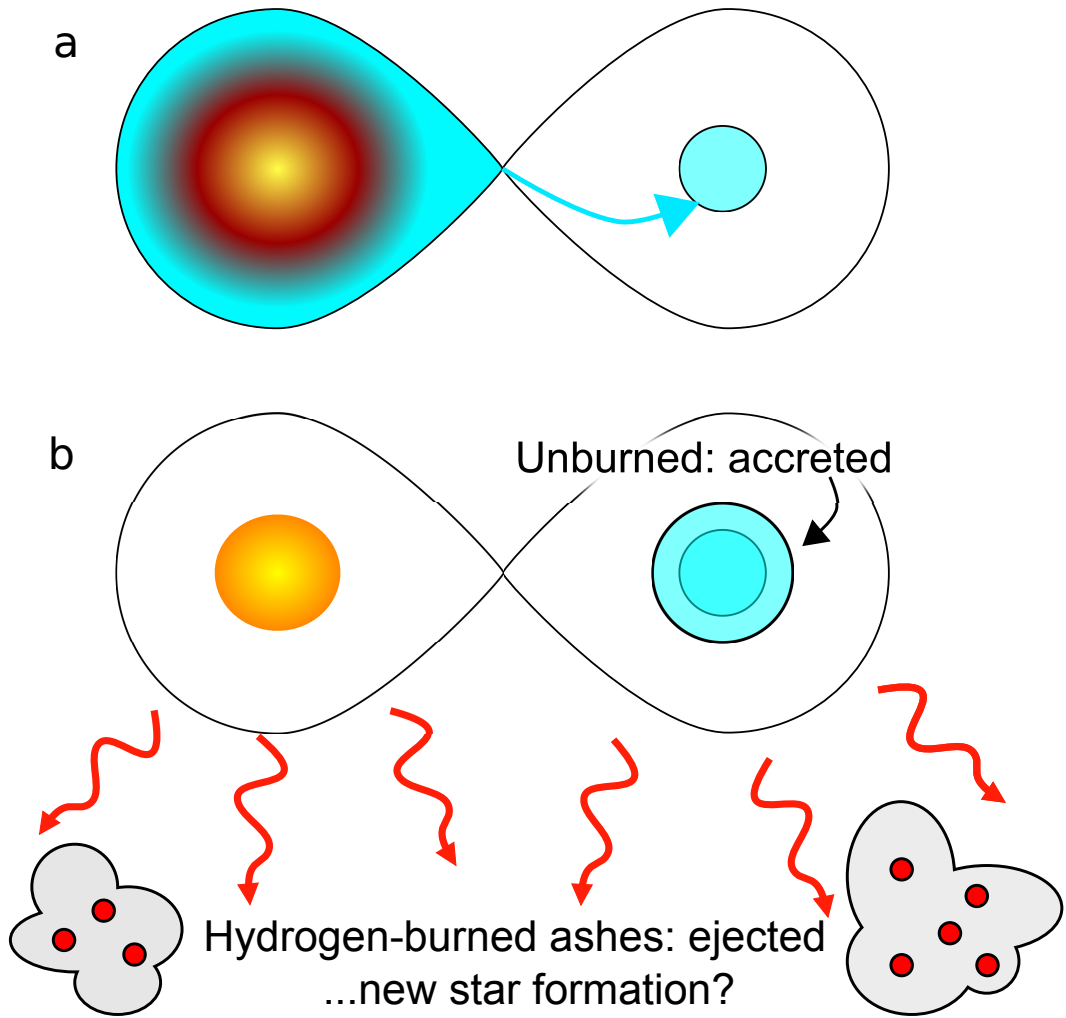}
\par\end{centering}

\caption{\label{fig:schematic-RLOF}Schematic view of Roche-lobe overflow in
a massive binary system. \textbf{(a)} At the start of Roche-lobe overflow,
the primary star (left) overflows its Roche lobe and transfers material
to the secondary (right). \textbf{(b)} By the end of Roche-lobe overflow,
the secondary has accreted unburned material while hydrogen-burned
material from deep inside the primary has been ejected from the binary
system and may mix with other sources of interstellar gas from which
a subsequent generation of stars may form.}

\end{figure*}

Because stars expand as they age, in a close binary the initially
more massive (primary) star overflows its Roche lobe first, transferring
mass onto the (initially less massive) secondary (Fig.~\ref{fig:schematic-RLOF}).
Material flows through the first Lagrange point onto the companion,
carrying with it both the chemical signature of the primary star and
angular momentum. The transferred mass settles onto the surface of
the secondary, spinning it up, but -- at least initially -- not greatly
altering its chemical abundance because material near the surface
of the primary is never hot enough for nuclear reactions to be efficient.

Accretion and spin up continues until the mass of the secondary increases
by about $10\%$, at which point it rotates so fast that material
at its equator is unbound \citep{1981A&A...102...17P}. Any further
mass transferred by Roche-lobe overflow is ejected from the binary
system at a velocity which is low compared to the proto-globular cluster
ejection speed. This material may be retained in the cluster for further
star formation. As the primary continues to transfer mass, it loses
its unburned envelope and material originally deep inside the star,
which has undergone nuclear burning, is exposed at the stellar surface.
First, layers burned by the CN cycle, then CNO, and later NeNa and
MgAl cycles, are transferred through the Lagrange point and ejected
from the binary system. Detailed binary evolution models suggest that
about three quarters of the transferred mass is ejected from a close
binary system, i.e. an accretion efficiency less than about $0.25$
\citep{2009A&A...507L...1D}, the binary-star physics remains highly
uncertain and its study continues \citep[e.g.][]{2011A&A...528A..16V,2012arXiv1211.3742D}.

While the binary-star scenario has not yet been explored in detail,
it is observed in nature. The binary star RY Scuti is ejecting material
rich in helium and nitrogen, and poor in oxygen and carbon, at a velocity
of about $50\,\mathrm{km}\,\mathrm{s}^{-1}$ \citep*{2001AJ....122.2700S}
i.e. more slowly than a stellar wind or the escape speed of a young
globular cluster. Further examples of binary mass transfer include
the Algol systems \citep{2011A&A...528A..16V}, X-ray binaries \citep{1977ApJ...212..533F}
and Wolf-Rayet binaries \citep{2005A&A...435.1013P} which must also
be products of non-conservative mass transfer. 

It is clear that a copious amount of material is ejected from interacting
binary stars, much of which has been processed by nuclear burning.
We estimate that as much as $13\%$ of the mass of a generation of
stars can be ejected in massive binaries, an amount similar to that
ejected from rapidly rotating massive stars and AGB stars combined
\citep{2009A&A...507L...1D}.

\section{Frascati-fuelled Perspective}

It is unlikely that anyone would bet more than a bottle of Frascati's
finest white wine on any single one of the proposed scenarios for
globular cluster self-pollution being the \emph{only} source of mass
for a second generation of stars. Massive AGB stars are generally
considered the best candidate because they can process material through
hot hydrogen-burning prior to its ejection in a slow wind, although
if third dredge up happens in these stars they may not be responsible
(although see \citealp{2008ApJ...684.1159Y}). The mass range which
contributes to clusters is unclear also, are super-AGB stars candidates
\citep{2012MNRAS.423.1521D}? Rapidly rotating massive stars certainly
exist, but their total ejected mass is not enough even assuming --
realistically? -- that they are all rapid rotators \citep{2009A&A...507L...1D}.
Binary stars may eject enough mass to satisfy the requirements of
a second stellar generation, but quite how conservative is binary
mass loss is not clear even after many decades of study \citep[e.g.][and references therein]{2007A&A...467.1181D}.
The competition between star formation and cluster gas ejection is
also relevant because massive stars evolve quickly relative to AGB
stars. It may be that massive-star ejecta escapes from the globular
cluster before forming any new stars (see e.g. Charbonnel et al. and
other contributions to this volume).

Uncertainties in stellar physics, e.g. mass-loss rates, mixing rates
and nuclear reaction rates, affect stellar yield predictions considerably
(e.g. \citealp{2005A&A...431..279V,2007A&A...466..641I,2007MNRAS.375.1280S,2009A&A...497..243D,2013arXiv1301.2487M};
and many others). The magnesium-aluminium negative correlation is
particularly difficult to reproduce because it requires proton capture
at temperatures which massive stars are unable to reach, while such
burning is possible in massive AGB stars \citep{2011MNRAS.415.3865V}.
Still, the massive-binary channel remains relatively unexplored and
a serious contributor to the mass that makes the second generation
of stars in globular clusters.
\begin{acknowledgements}
RGI thanks the conference organisers, the Alexander von Humboldt Foundation
for supporting his work and Richard Stancliffe for a critical reading
of the manuscript.\vspace{-3mm}
\end{acknowledgements}
\bibliographystyle{aa}
%\bibliography{/vol/aibn36/aibn36_1/izzard/svn/tex/references}

\begin{thebibliography}{28}
\expandafter\ifx\csname natexlab\endcsname\relax\def\natexlab#1{#1}\fi

\bibitem[{{Carretta} {et~al.}(2009){Carretta}, {Bragaglia}, {Gratton},
  {Lucatello}, {Catanzaro}, {Leone}, {Bellazzini}, {Claudi}, {D'Orazi},
  {Momany}, {Ortolani}, {Pancino}, {Piotto}, {Recio-Blanco}, \&
  {Sabbi}}]{2009A&A...505..117C}
{Carretta}, E., {Bragaglia}, A., {Gratton}, R.~G., {et~al.} 2009, \aap, 505,
  117

\bibitem[{{de Mink} {et~al.}(2009{\natexlab{a}}){de Mink}, {Cantiello},
  {Langer}, {Pols}, {Brott}, \& {Yoon}}]{2009A&A...497..243D}
{de Mink}, S.~E., {Cantiello}, M., {Langer}, N., {et~al.} 2009{\natexlab{a}},
  \aap, 497, 243

\bibitem[{{de Mink} {et~al.}(2013){de Mink}, {Langer}, {Izzard}, {Sana}, \& {de
  Koter}}]{2012arXiv1211.3742D}
{de Mink}, S.~E., {Langer}, N., {Izzard}, R.~G., {Sana}, H., \& {de Koter}, A.
  2013, \apj \hspace{2mm}(in press), ArXiv 1211.3742

\bibitem[{{de Mink} {et~al.}(2007){de Mink}, {Pols}, \&
  {Hilditch}}]{2007A&A...467.1181D}
{de Mink}, S.~E., {Pols}, O.~R., \& {Hilditch}, R.~W. 2007, \aap, 467, 1181

\bibitem[{{de Mink} {et~al.}(2009{\natexlab{b}}){de Mink}, {Pols}, {Langer}, \&
  {Izzard}}]{2009A&A...507L...1D}
{de Mink}, S.~E., {Pols}, O.~R., {Langer}, N., \& {Izzard}, R.~G.
  2009{\natexlab{b}}, \aap, 507, L1

\bibitem[{{Decressin} {et~al.}(2007){Decressin}, {Meynet}, {Charbonnel},
  {Prantzos}, \& {Ekstr{\"o}m}}]{2007A&A...464.1029D}
{Decressin}, T., {Meynet}, G., {Charbonnel}, C., {Prantzos}, N., \&
  {Ekstr{\"o}m}, S. 2007, \aap, 464, 1029

\bibitem[{{D'Ercole} {et~al.}(2012){D'Ercole}, {D'Antona}, {Carini},
  {Vesperini}, \& {Ventura}}]{2012MNRAS.423.1521D}
{D'Ercole}, A., {D'Antona}, F., {Carini}, R., {Vesperini}, E., \& {Ventura}, P.
  2012, \mnras, 423, 1521

\bibitem[{{Flannery} \& {Ulrich}(1977)}]{1977ApJ...212..533F}
{Flannery}, B.~P. \& {Ulrich}, R.~K. 1977, \apj, 212, 533

\bibitem[{{Fuhrmann} \& {Chini}(2012)}]{2012ApJS..203...30F}
{Fuhrmann}, K. \& {Chini}, R. 2012, \apjs, 203, 30

\bibitem[{{Glebbeek} {et~al.}(2009){Glebbeek}, {Gaburov}, {de Mink}, {Pols}, \&
  {Portegies Zwart}}]{2009A&A...497..255G}
{Glebbeek}, E., {Gaburov}, E., {de Mink}, S.~E., {Pols}, O.~R., \& {Portegies
  Zwart}, S.~F. 2009, \aap, 497, 255

\bibitem[{{Izzard}(2004)}]{2004MmSAI..75..754I}
{Izzard}, R.~G. 2004, Memorie della Societa Astronomica Italiana, 75, 754

\bibitem[{{Izzard} {et~al.}(2007){Izzard}, {Lugaro}, {Karakas}, {Iliadis}, \&
  {van Raai}}]{2007A&A...466..641I}
{Izzard}, R.~G., {Lugaro}, M., {Karakas}, A.~I., {Iliadis}, C., \& {van Raai},
  M. 2007, \aap, 466, 641

\bibitem[{{Kouwenhoven} {et~al.}(2007){Kouwenhoven}, {Brown}, {Portegies
  Zwart}, \& {Kaper}}]{2007A&A...474...77K}
{Kouwenhoven}, M.~B.~N., {Brown}, A.~G.~A., {Portegies Zwart}, S.~F., \&
  {Kaper}, L. 2007, \aap, 474, 77

\bibitem[{{Meynet} {et~al.}(2013){Meynet}, {Ekstr{\"o}m}, {Maeder},
  {Eggenberger}, {Saio}, {Chomienne}, \& {Haemmerl{\'e}}}]{2013arXiv1301.2487M}
{Meynet}, G., {Ekstr{\"o}m}, S., {Maeder}, A., {et~al.} 2013, ArXiv e-print
  1301.2487

\bibitem[{{Packet}(1981)}]{1981A&A...102...17P}
{Packet}, W. 1981, \aap, 102, 17

\bibitem[{{Petrovic} {et~al.}(2005){Petrovic}, {Langer}, \& {van der
  Hucht}}]{2005A&A...435.1013P}
{Petrovic}, J., {Langer}, N., \& {van der Hucht}, K.~A. 2005, \aap, 435, 1013

\bibitem[{{Prantzos} {et~al.}(2007){Prantzos}, {Charbonnel}, \&
  {Iliadis}}]{2007A&A...470..179P}
{Prantzos}, N., {Charbonnel}, C., \& {Iliadis}, C. 2007, \aap, 470, 179

\bibitem[{{Raghavan} {et~al.}(2010){Raghavan}, {McAlister}, {Henry}, {Latham},
  {Marcy}, {Mason}, {Gies}, {White}, \& {ten Brummelaar}}]{2010ApJS..190....1R}
{Raghavan}, D., {McAlister}, H.~A., {Henry}, T.~J., {et~al.} 2010, \apjs, 190,
  1

\bibitem[{{Sana} {et~al.}(2012){Sana}, {de Mink}, {de Koter}, {Langer},
  {Evans}, {Gieles}, {Gosset}, {Izzard}, {Le Bouquin}, \&
  {Schneider}}]{2012Sci...337..444S}
{Sana}, H., {de Mink}, S.~E., {de Koter}, A., {et~al.} 2012, Science, 337, 444

\bibitem[{{Sills} \& {Glebbeek}(2010)}]{2010MNRAS.407..277S}
{Sills}, A. \& {Glebbeek}, E. 2010, \mnras, 407, 277

\bibitem[{{Smith} {et~al.}(2001){Smith}, {Gehrz}, \&
  {Goss}}]{2001AJ....122.2700S}
{Smith}, N., {Gehrz}, R.~D., \& {Goss}, W.~M. 2001, \aj, 122, 2700

\bibitem[{{Stancliffe} \& {Jeffery}(2007)}]{2007MNRAS.375.1280S}
{Stancliffe}, R.~J. \& {Jeffery}, C.~S. 2007, MNRAS, 375, 1280

\bibitem[{{van Rensbergen} {et~al.}(2011){van Rensbergen}, {de Greve},
  {Mennekens}, {Jansen}, \& {de Loore}}]{2011A&A...528A..16V}
{van Rensbergen}, W., {de Greve}, J.~P., {Mennekens}, N., {Jansen}, K., \& {de
  Loore}, C. 2011, \aap, 528, A16

\bibitem[{{Vanbeveren} {et~al.}(2012){Vanbeveren}, {Mennekens}, \& {De
  Greve}}]{2012A&A...543A...4V}
{Vanbeveren}, D., {Mennekens}, N., \& {De Greve}, J.~P. 2012, \aap, 543, A4

\bibitem[{{Ventura} {et~al.}(2011){Ventura}, {Carini}, \&
  {D'Antona}}]{2011MNRAS.415.3865V}
{Ventura}, P., {Carini}, R., \& {D'Antona}, F. 2011, \mnras, 415, 3865

\bibitem[{{Ventura} \& {D'Antona}(2005)}]{2005A&A...431..279V}
{Ventura}, P. \& {D'Antona}, F. 2005, \aap, 431, 279

\bibitem[{{Ventura} {et~al.}(2001){Ventura}, {D'Antona}, {Mazzitelli}, \&
  {Gratton}}]{2001ApJ...550L..65V}
{Ventura}, P., {D'Antona}, F., {Mazzitelli}, I., \& {Gratton}, R. 2001, \apjl,
  550, L65

\bibitem[{{Yong} {et~al.}(2008){Yong}, {Grundahl}, {Johnson}, \&
  {Asplund}}]{2008ApJ...684.1159Y}
{Yong}, D., {Grundahl}, F., {Johnson}, J.~A., \& {Asplund}, M. 2008, \apj, 684,
  1159

\end{thebibliography}

\end{document}